\documentclass[a4paper,11pt]{article}
\usepackage{pos}
\usepackage{subcaption}
\usepackage{enumitem}

\title{Vertically stacked amorphous selenium based VUV photodetectors for use in liquid noble detectors}

\author*[a]{Iakovos Tzoka}
\author[a,b]{M. Rooks}
\author[a]{A. C. A. Ishida}
\author[a]{A. Barajas}
\author[a]{V. A. Chirayath}
\author[a]{J. Asaadi}

\affiliation[a]{University of Texas at Arlington,\\
  701 S. Nedderman Drive, Arlington, TX, 76019, USA}
\affiliation[b]{Oak Ridge National Laboratory,\\
  5200, 1 Bethel Valley Rd, Oak Ridge, TN 37830}

\emailAdd{iakovos.tzoka@uta.edu}

\abstract{We present results from the characterization of a vertically stacked amorphous selenium (aSe)-based photodetector for use in cryogenic environments. aSe has been identified as an ideal photoconductor that can efficiently convert vacuum ultraviolet (VUV) light to charges even at cryogenic temperatures. We have designed and fabricated an aSe device in vertical geometry with top and bottom metal electrodes that produces an electric field perpendicular to the substrate. The top-metal contact has an open design that results in a large fraction of the aSe thin film surface to be active for photodetection. Our experiments show that the vertically stacked aSe device detects light from a Xenon flash lamp in a vacuum environment and can produce measurable signals at \(\sim \)130K. We also demonstrate a significant enhancement in the amplitude of the photoinduced signal by growing graphene on the top-metal contact and the aSe thin film. Our results provide the first demonstration of a vertical aSe based VUV photodetector that utilizes the wide-band optical transparency of graphene top-electrode. Our results could open the doorway to a potentially game-changing solution of an integrated charge and light sensor that can be employed in future large-scale time projection chambers with pixelated anode planes.}

\FullConference{%
   ICHEP 2024 \\
   07/18/2024 \\
   Prague Conference Center, Prague, Czech Republic 
}

\begin{document}
\maketitle

\section{Introduction}
Liquid argon (LAr) and xenon (LXe), two commonly used noble elements in time projection chambers (TPCs), act as both the detection medium and scintillation medium, offering high scintillation yields in the vacuum ultraviolet (VUV) range. However, detecting VUV photons with common photosensors such as SiPMs and PMTs is difficult, leading to the use of wavelength-shifting materials like tetraphenyl butadiene and polyethylene naphthalate, which shift the scintillation light into the longer wavelength region \cite{Ku_niak_2020}. However, these materials have low conversion efficiencies \cite{Benson_2018} and deteriorate quickly under harsh environmental conditions, such as cryogenic temperatures \cite{Abraham_2021}. This has led to research into alternative materials, such as amorphous selenium (aSe), which has been well studied for applications in X-ray imaging \cite{James_R_2015, 8886491}. Due to its unique electronic structure, aSe can efficiently convert VUV photons into electron-hole pairs \cite{Barman}. Its good effective hole mobility and the potential for impact ionization-induced hole multiplication at large applied fields make aSe an excellent candidate for detecting events with low photon yields at cryogenic temperatures.

Rooks et al. \cite{Rooks:2022vrl} recently demonstrated, using a horizontal device geometry, that an aSe-based device remains robust under cryogenic conditions and produces a response at temperatures relevant to liquid noble detectors commonly used in high-energy physics. This demonstration by our group, using a novel windowless horizontal geometry with aSe coated over interdigitated metal electrodes, provides a pathway to implementing a fully pixelated anode plane capable of detecting both charge and light. However, the electric field and the subsequent carrier transport in aSe, achieved with the interdigitated electrodes in horizontal geometry, are non-uniform and non-trivial to investigate, making full optimization of the detector difficult. Additionally, the side-by-side arrangement of interdigitated metal electrodes necessitates that the hole-blocking layer, essential for operation in the avalanche regime, be selectively coated only on the anode to avoid carrier-induced heating of the hole-blocking layer if it is also present on the cathode.

Here, we explore an aSe-based device with a vertical geometry, i.e., aSe placed between two charge-collecting conductors, but with a top metal electrode design that allows for a large aSe active region. We compare the signal obtained from this novel detector to one where a VUV-transparent graphene \cite{Zheng} top electrode is grown on aSe to make the electric field in aSe more uniform.

\section{Device Preparation}
\begin{figure}[htbp]
    \centering
    \begin{subfigure}[b]{0.35\textwidth}
        \centering
        \includegraphics[width=\textwidth]{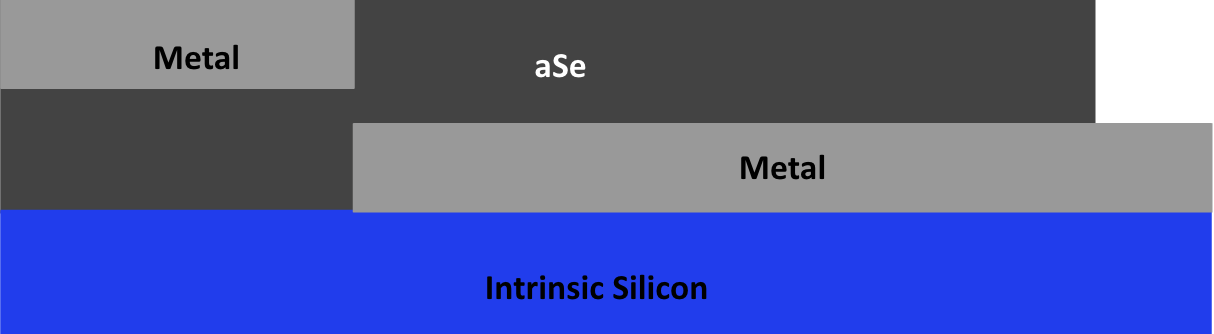}
        \caption{Cross sectional view of the vertically stacked aSe device before graphene deposition.}
        \label{fig:1a}
    \end{subfigure}
    \begin{subfigure}[b]{0.3\textwidth}
        \centering 
        \includegraphics[width=\textwidth]{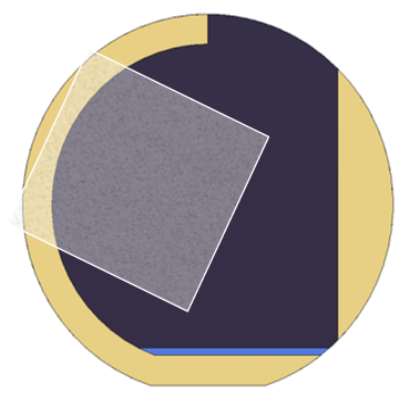}
        \caption{Schematic representation of the vertically stacked aSe device after graphene growth.}
        \label{fig:1b}
    \end{subfigure}    
    \caption{aSe device geometry} 
    \label{fig:devicefabtotal}
\end{figure}

The aSe devices were grown on intrinsic silicon with a large surface resistance (>10kohm/cm). The 2-inch double-sided, polished (DSP) silicon wafers with a thickness of $500\pm25\mu$m were sourced from WaferPro \cite{waferpro}. Prior to metal growth, the wafers were cleaned using a mixture of 100 mL of sulfuric acid (99\% concentration) and hydrogen peroxide (30\% concentration) to remove any organic contaminants. The  first metal contact is grown on the wafer using DC magnetron sputtering. We have used two types of metal contact. The devices shown here has either $1\mu$m thick chromium contact or Titanium (990nm)/Gold (10 nm) (Ti/Au) metal contact. A $1\mu$m thick aSe is thermally evaporated on the metal contact and the substrate using high purity selenium pellets stabilized with 0.2\% arsenic and 10 ppm of chlorine.  In the next stage, an electron beam thermal evaporator is employed to deposit the top metal contact (chrome or Ti/Au) on to the aSe film. The electron beam evaporator was chosen to prevent crystallization in the aSe. The large throw distance possible in the electron beam evaporation system keeps the aSe film below the temperatures at which crystallization occurs. A cross-sectional view of the device after the top metal contact growth is shown in Figure \ref{fig:1a}. These devices are tested by exposing light from Xenon flash lamp under vacuum at room temperature and at cryogenic temperatures. The top electrode is extended by depositing multilayer graphene using the wet transfer method as instructed by the supplier ACS materials \cite{ACS}. A schematic of the completed vertical device after the graphene deposition is shown in Figure \ref{fig:1b}.

\section{Experimental Setup}
\begin{figure}[htbp]
    \centering
    \begin{subfigure}[b]{0.3\textwidth}
        \centering
        \includegraphics[width=\textwidth]{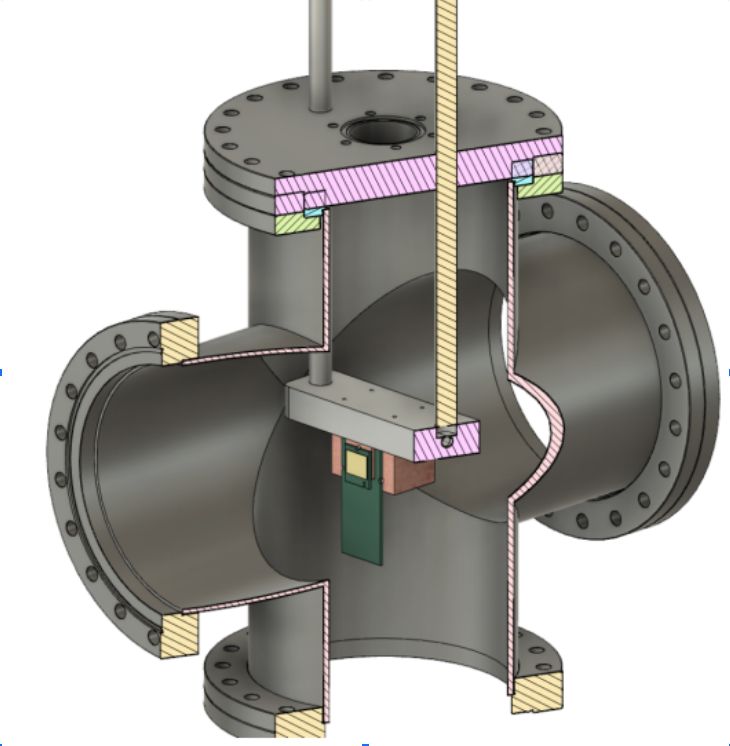}
        \caption{High vacuum chamber and the liquid nitrogen bath for controlling the device temperature}
        \label{fig:2a}
    \end{subfigure}
    \begin{subfigure}[b]{0.3\textwidth}
        \centering
        \includegraphics[width=\textwidth]{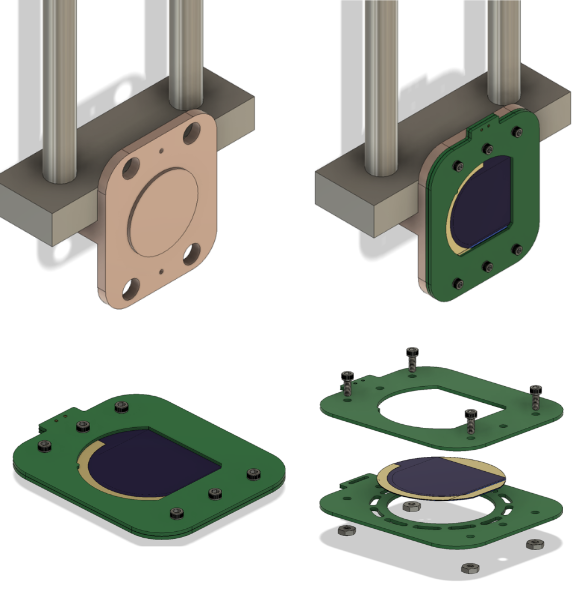}
        \caption{Copper holder sitting on the liquid nitrogen bath and the PCB Holder for Vertical device}
        \label{fig:2b}
    \end{subfigure}
    \caption{Device testing system} 
    \label{fig:mainfigure} 
\end{figure}

The experimental setup used to test these devices in vacuum (Figure \ref{fig:2a}) is described in  \cite{Rooks:2022vrl}. The setup was modified to receive a device in the vertical geometry. The aSe devices are mounted between two PCBs in a sandwich geometry to make contact with the top and bottom metal electrodes. The PCB device holder is mounted on a copper holder such that the substrate (Si wafer) touches the copper providing a good thermal contact. The copper holder is mounted on a cold bath which consists of a stainless-steel bar through which liquid nitrogen can flow. The 3D CAD drawing of the device mounting method is shown in Figure \ref{fig:2b}.  

We have used an HAMAMATSU 5W Xenon flash lamp \cite{Hamamatsu} as the light source for testing the devices as they have a wide emission spectrum from 200 nm to 1000 nm. The flash lamp is pulsed at the desired frequency using the RIGOL DG1062Z\cite{Rigol} arbitrary waveform generator. For the purposes of testing, the Xenon flash lamp is pulsed at 1 Hz with a duty cycle of 1\%. The light is coupled to the vacuum system using a VUV-compatible optic fiber. The device is biased to the desired voltage using a custom-built-high-voltage DC power supply. The capacitively decoupled signal is fed into a A250 charge sensitive pre-amplifier from AMETEK \cite{AMETEK}. The data is acquired using a LeCroy WaveRunner 677A oscilloscope. We acquire the signals both before and after the amplification process. 

We controlled the area of exposure and the amount of light on the device by masking the active area. One of the masks exposes the full active area of $12.5 \text{cm}^2$ of the amorphous selenium, while the smaller mask  exposes $4 \text{cm}^2$ of the total area of the amorphous selenium.

\section{Results}
\begin{figure}[htbp]
    \centering
    \includegraphics[width=0.5\textwidth]{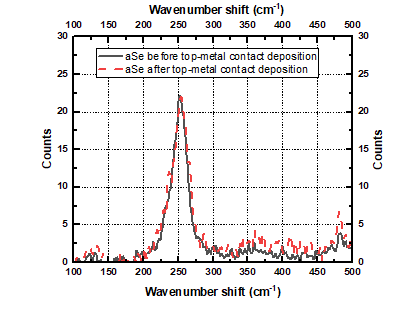}
    \caption{Raman spectrum obtained from the aSe device before and after top-metal deposition. The peak at 250 $cm^{-1}$ is broad and devoid of features seen for m-Se or t-Se \cite{Goldan} pointing to amorphous structure. The peak does not have a change in shape after metal deposition.}
    \label{fig:3}
\end{figure}

We used Raman spectroscopy to confirm that the top-metal deposition does not affect the amorphous structure of the underlying selenium. Raman spectrum (Figure \ref{fig:3}) obtained from aSe before and after metal deposition shows that the peak at 250 $cm^{-1}$ remains unchanged and confirms that the amorphous selenium is unaffected.

Before depositing graphene on aSe thin film and the top-metal electrode, the vertical device is tested by illuminating the device with a Xenon flash lamp at different applied voltages. Our results show that:  
\begin{enumerate}[noitemsep,nolistsep]
    \item the peak amplitude of the photoinduced signal increases with the increase of the applied voltage across the aSe thin film,
    \item the polarity of the photoinduced signal at applied voltage depends on the nature of the metal used and
    \item the peak amplitude of the photoinduced signal decreases with decreasing temperature. 
\end{enumerate}

\begin{figure}[htbp]
    \centering
    \begin{subfigure}[b]{0.4\textwidth}
        \centering
        \includegraphics[width=\textwidth]{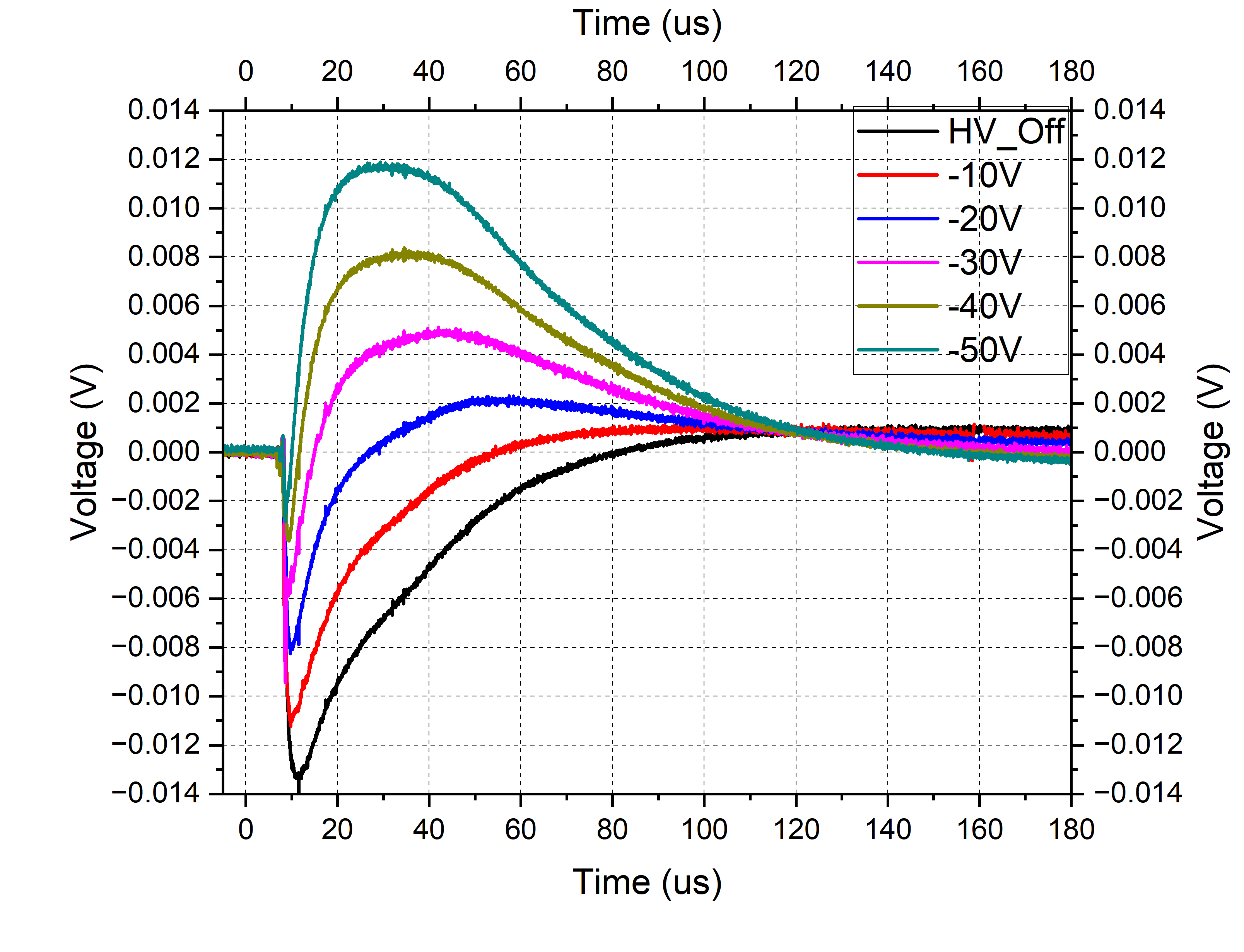}
        \caption{The photoinduced signals from 1$\mu$m aSe devices becomes more positive with increasing negative bias on the bottom Ti/Au contacts.}
        \label{fig:Figure 4a}
    \end{subfigure}
    \begin{subfigure}[b]{0.4\textwidth}
        \centering
        \includegraphics[width=\textwidth]{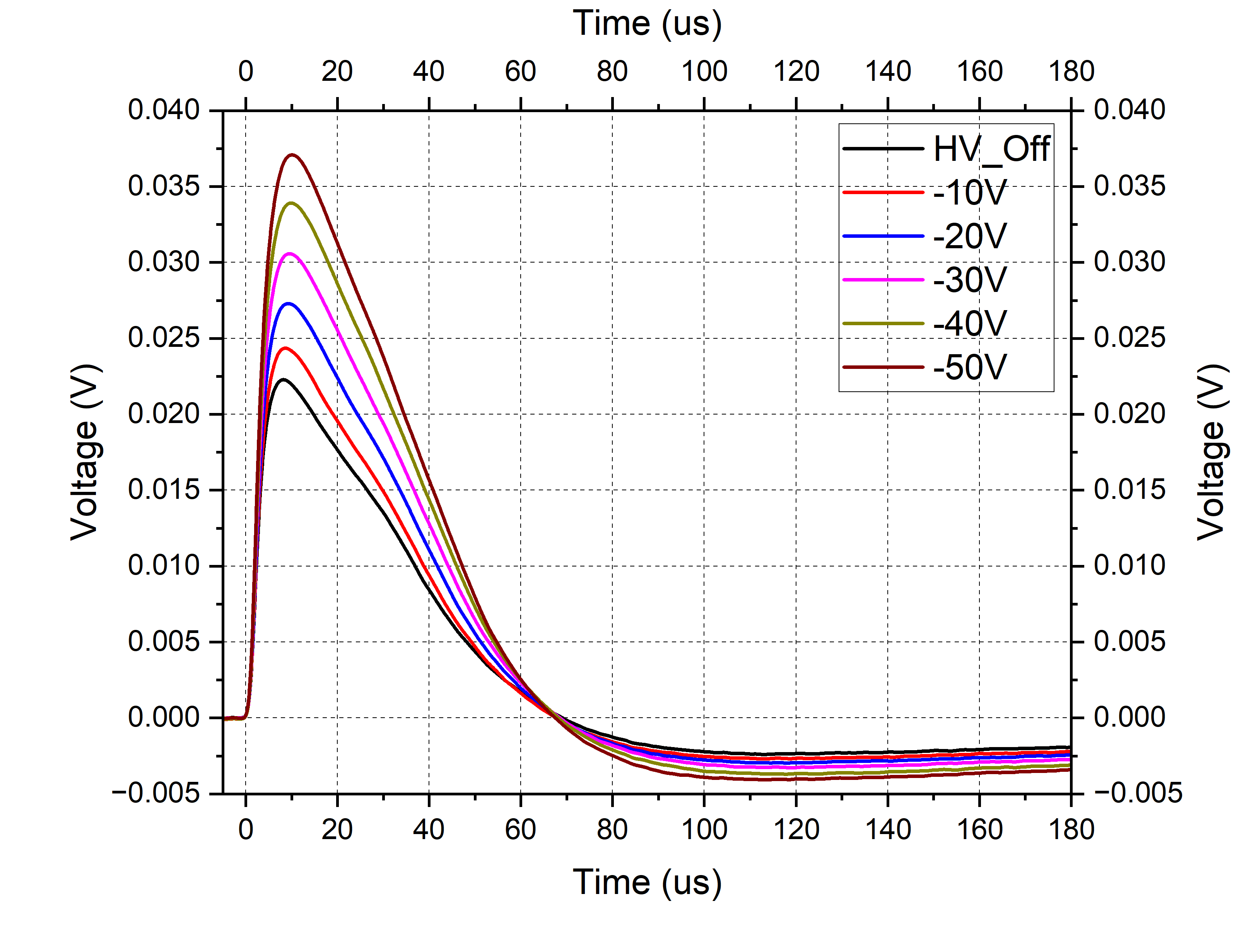}
        \caption{The amplitude of the photoinduced signals from 1$\mu$m aSe devices increases with the applied negative bias on the bottom Cr contact.}
        \label{fig:Figure 4b}
    \end{subfigure}
    \caption{Variation of the photoinduced signal with applied potential} 
    \label{fig:} 
\end{figure}

The increase in the amplitude of the photoinduced signal with the applied voltage is shown in Figure \ref{fig:Figure 4a} and Figure \ref{fig:Figure 4b}. For devices with Ti/Au contacts (Figure \ref{fig:Figure 4a}), the photoinduced signal at zero bias has a negative polarity. As the negative potential on the bottom electrode increased, the photoinduced signal became more positive. For devices with chromium metal contact, the photoinduced signals collected at the bottom electrode with zero bias were positive. The amplitude of the positive going signal increased with the negative potential on the bottom electrode. In both cases, an increasing negative potential on the bottom metal contact resulted in an increase in the amplitude pointing to an increase in the number of holes collected at the bottom electrode. These results were obtained with full mask that exposes the full active region of the aSe detector and are non-amplified. 

Please note that the metal used for making the contact determines the direction of the inherent electric field at the metal-aSe interface.  The aSe devices with Ti/Au contact resulted in a negative going signal at zero bias whereas the devices with Cr contact resulted in positive signals. In the case of Ti/Au, the inherent electric field at the metal-aSe interface favors collection of electrons and thus, produces a negative signal. While for Cr, the hole collection is favored resulting in a positive signal when no potential is applied across the aSe thin film. 

\begin{figure}[htbp]
    \centering
    \begin{subfigure}[b]{0.4\textwidth}
        \centering
        \includegraphics[width=\textwidth]{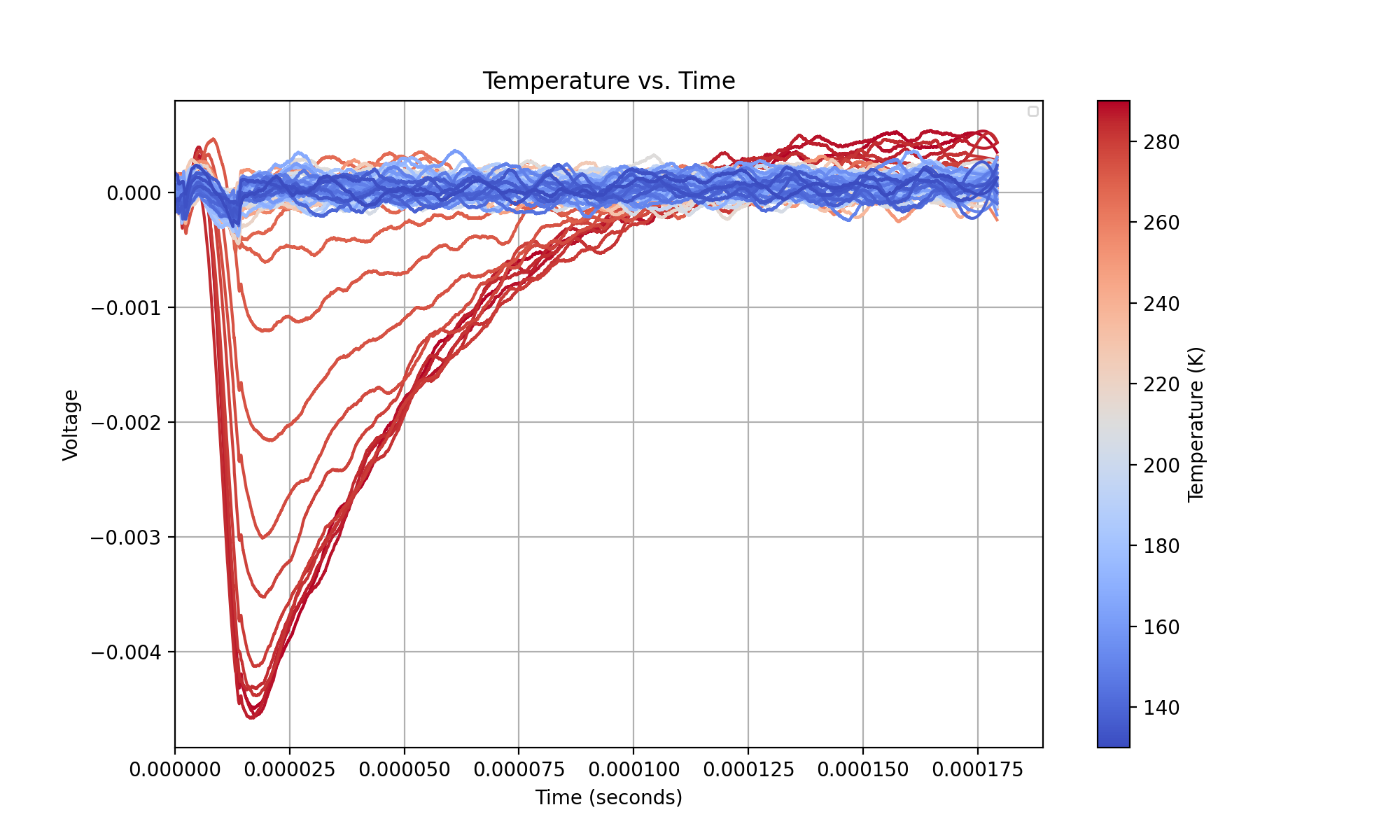}
        \caption{The amplitude of the photoinduced signals reduces as a function of temperature from 293K to 125K.}
        \label{fig:5a}
    \end{subfigure}
    \begin{subfigure}[b]{0.42\textwidth}
        \centering
        \includegraphics[width=\textwidth]{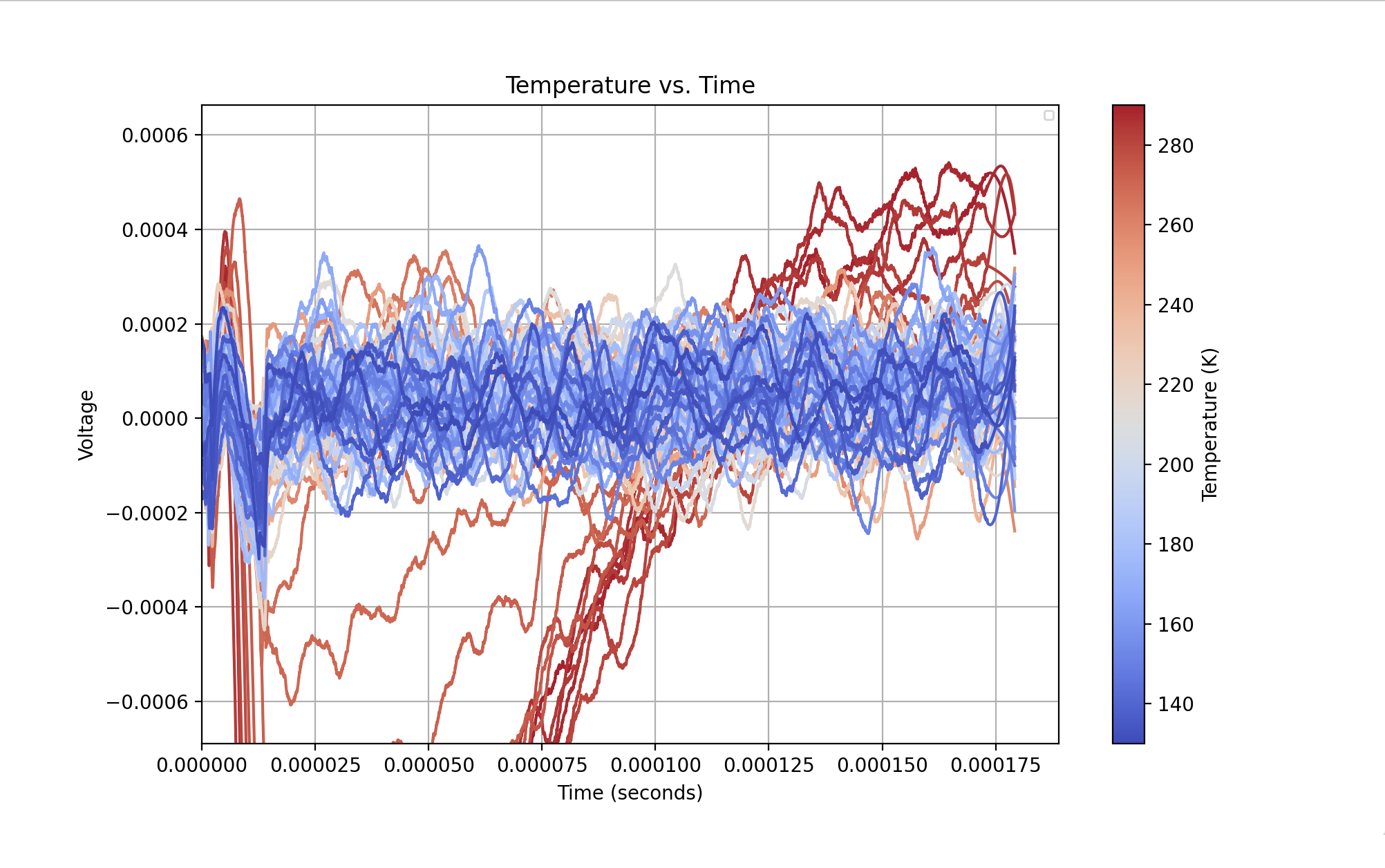}
        \caption{The zoomed in version to show the existence of the signal even in cryogenic temperatures.}
        \label{fig:5b}
    \end{subfigure}
    \caption{Testing of the vertically stacked aSe device as a function of temperature}
    \label{fig:Fig. 5}
\end{figure}

The testing of the vertically stacked 1$\mu$m aSe device with Ti/Au metal contact as a function of temperature is shown in Figure \ref{fig:5a}. Please note that these measurements were done with the Xe flash lamp flashing at 25 mHz. One photoinduced pulse is collected every 40 seconds for which the temperature is known and is shown in the color map. The signals were collected at an applied bias of -30V on the bottom metal electrode. It can be seen that the peak amplitude has decreased with the temperature and a small photoinduced signal was still visible at \(\sim \) 130K(Figure \ref{fig:5b}). The reduction in intensity with temperature is attributed to the lower collection efficiency due to the reduced mobilities of the charge carriers and the reduced probability for detrapping from defect states at lower temperatures. 

\begin{figure}
    \centering
    \includegraphics[width=0.4\linewidth]{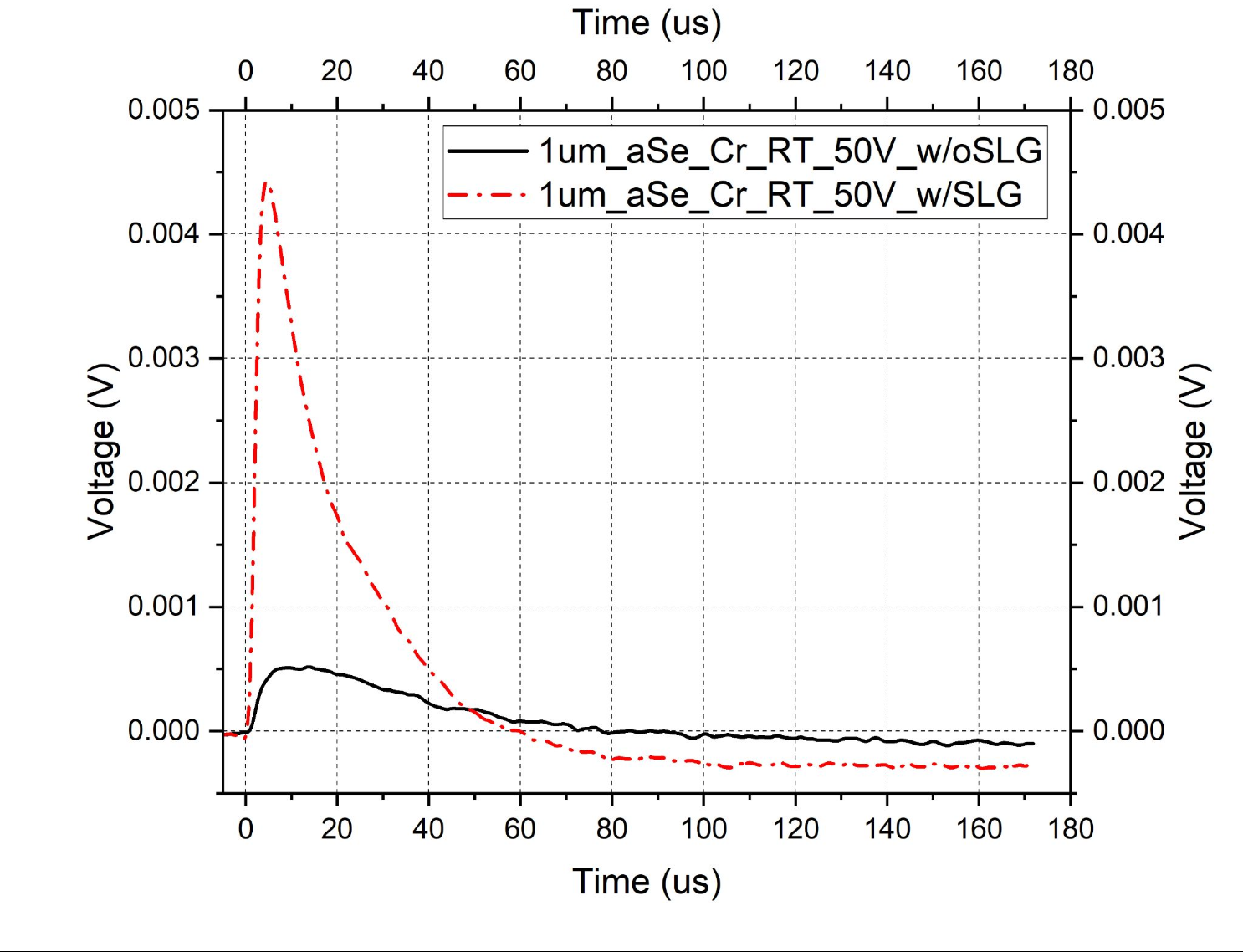}
    \caption{Comparison of the same sample with and without graphene. Peak amplitude increased from 0.5 mV to 45mv approximately an order of 10.}
    \label{fig:Fig. 6}
\end{figure}

We increased the surface area of the top metal electrode for the aSe device with Cr metal contact by depositing growing multilayer graphene. The device performance before the deposition of graphene is compared with that after graphene deposition in Figure \ref{fig:Fig. 6}. Please note that for these measurements we used a smaller mask for illuminating only the region of the device where graphene was grown. We saw a significant increase in the peak amplitude (by a factor of ten) with graphene deposition. The increase is understood to be due to the more uniform electric field and better charge collection as a result of it. 

\section{Conclusion}

 We have demonstrated the functionality of a novel vertically stacked aSe device utilizing graphene as the top electrode. The presence of graphene increased the uniformity of the electric field across the aSe that resulted in increased charge collection efficiency and thus larger signal amplitudes. We also demonstrated that the vertically stacked aSe devices showed a measurable signal even without the top graphene electrode. We will extend our investigations to test performance of the aSe-graphene devices at cryogenic temperatures. We will remake the devices on fused silica substrates instead of silicon, which should allow us to reach higher applied fields. We also aim to investigate the suitability of different metal contacts and their behavior as a function of temperature. 

 \section{Acknowledgements} The authors would like to thank the Center of Advanced Detector Technologies at UTA that gave us the opportunity to present our results at ICHEP. VAC acknowledge the support of NSF Grant No. CHE – 2204230. JA acknowledges support from awards DE-SC0020065 and DE-SC0000253485. MR acknowledges support from award FNAL-LDRD-2020-027.



\bibliography{skeleton} {}
\bibliographystyle{ieeetr}

\end{document}